\documentstyle[12pt,psfig]{article}
\newcommand{\beq}{\begin{equation}}
\newcommand{\eeq}{\end{equation}}
\newcommand{\bea}{\begin{eqnarray}}
\newcommand{\eea}{\end{eqnarray}}

\renewcommand{\d}{\delta}

\newcommand{\oh}{\frac{1}{2}}

\newcommand{\dg}{\dagger}
\newcommand{\non}{\nonumber}

\begin{document}

\addtolength{\baselineskip}{0.20\baselineskip}

\hfill hep-lat/9611002

\hfill LBNL-39472

\hfill October 1996

\begin{center}

\vspace{24pt}

{\large \bf The Abelianicity of Cooled $SU(2)$ Lattice Configurations }

\end{center}

\vspace{12pt}

\begin{center}

{\sl J. Giedt${}^a$} and {\sl J. Greensite${}^{a,b}$}

\end{center}

\vspace{12pt}

\begin{tabbing}

{}~~~~~~~~~~~~~\= blah  \kill
\> ${}^a$ Physics and Astronomy Dept., San Francisco State University, \\
\> ~~San Francisco CA 94132 USA.  E-mail: {\tt jgiedt@stars.sfsu.edu}, \\
\> ~~{\tt greensit@stars.sfsu.edu} \\
\\
\> ${}^b$ Theoretical Physics Group, Ernest Orlando Lawrence Berkeley \\
\> ~~National Laboratory, University of California, Berkeley, \\
\> ~~California 94720 USA.  E-mail: {\tt greensite@theorm.lbl.gov} \\

\end{tabbing}

\vspace{18pt}

\begin{center}

{\bf Abstract}

\end{center}

\bigskip

    We introduce a gauge-invariant measure of the local "abelianicity" of 
any given lattice configuration in non-abelian lattice gauge theory; it
is essentially a comparison of the magnitude of field strength commutators
to the magnitude of the field strength itself. 
This measure, in conjunction with the cooling technique, is used to
probe the $SU(2)$ lattice vacuum for a possible large-scale abelian 
background, underlying the local short-range field fluctuations.
We do, in fact, find a substantial rise in abelianicity over 10 cooling 
steps or so, after which the abelianicity tends to drop again.

\vfill

\newpage

    It has been suggested on occasion that non-abelian gauge
theory is dominated, in the infrared regime, by abelian configurations
of some kind.  One early example was Saviddy's proposal \cite{Sav}, based on a 
study of the one-loop effective action, that there is a constant background 
abelian field strength in the Yang-Mills vacuum.  More recently, in the
maximal abelian gauge, it has been shown that $SU(2)$ lattice configurations
that are "abelian projected" onto $U(1)$ configurations retain information
about the asymptotic string tension \cite{Suz}.  This property is known as 
"abelian dominance," and is widely interpreted as supporting 't Hooft's
abelian projection theory of confinement \cite{tH1} (see, however, ref.
\cite{Lat96}).  Some other ideas 
concerning the non-perturbative dynamics have stressed the importance 
of vortices carrying magnetic flux associated with the center elements of the
gauge group \cite{Z2}.  The center subgroup is, of course, abelian by 
definition.

    With this motivation it is interesting to study, via lattice
Monte Carlo simulations, the actual degree of "abelianicity" of vacuum 
fluctuations in non-abelian gauge theories. In order to do this, we must first
introduce a quantitative, and preferably gauge-invariant, measure of the 
abelianicity of a gauge-field configuration.  A non-abelian field
configuration may be regarded as equivalent to an abelian field
if the commutators of its field-strength components vanish everywhere; 
thus the following gauge-invariant, positive
semidefinite quantity
\beq
      B = -{1 \over V} \sum_x {1 \over n_p (n_p - 1)} 
   \sum_{i>j} \sum_{m>n} \mbox{Tr} \{ [F_{ij}(x),F_{mn}(x)]^2 \}
\label{A}
\eeq
vanishes if and only if the configuration is abelian.  On the lattice,
$V$ is the number of lattice sites, $n_p=D(D-1)/2$ is the number
of plaquettes per site in $D$ dimensions, with lattice field 
strength taken to be
\beq
  F_{ij} = {1\over 2i} [ U_i(x) U_j(x+i) U^\dg_i(x+j) U^\dg_j(x)
                        ~~ - ~~ \mbox{h.c.} ]
\eeq    
Of course, since $B$ is proportional to the fourth power of 
field-strengths, it is sensitive not only to abelianicity but also
to the magnitude of the field strengths.  For the purpose of
normalization, we introduce
\beq
       A = {1 \over n_p V} \sum_x \sum_{i>j} \mbox{Tr}\{ F_{ij}^4 \}
\label{B}
\eeq
and define the average {\it non}-abelianicity of an ensemble of 
configurations, which is invariant under a rescaling of the field-strengths, 
as
\beq
       Q \equiv {<B> \over <A>}
\eeq
An ensemble
of configurations is abelian if the non-abelianicity $Q$ vanishes.
This measure of non-abelianicity is also a quantitative measure of 
abelianicity, in the sense that abelianicity increases as $Q$ decreases.

   The observable $Q$ is a local quantity, and will be dominated by 
high-frequency vacuum fluctuations.\footnote{In zeroth-order lattice
perturbation theory, $Q=1.6$.}  
However, we are not so much interested in these
high-frequency fluctuations, which are perturbative in character, as
in the degree of abelianicity of the underlying larger-scale fluctuations.
One method that has been suggested for eliminating the higher-frequency
fluctuations from a given configuration is the lattice "cooling"
technique.  We have therefore measured the variation of $Q$ with 
cooling step, to determine whether the larger scale vacuum fluctuations are
more (or less) abelian in character than the high frequency fluctuations.

   We follow the constrained cooling technique of ref. \cite{Pisa}.
Each cooling step is a sweep through the lattice links, with each
link $U_k(x)$ replaced by a new link $U'_k(x)$ which minimizes the 
lattice action, subject to the constraint
\beq
         \oh  \mbox{Tr}[(U'_k(x)-U_k(x))
              (U'^\dagger_k(x)-U^\dagger_k(x))] \le \d^2
\eeq
where $\d=0.05$.
As pointed out by Teper \cite{Teper}, this (or any other) version of
cooling can never remove confinement entirely, because cooling
can only remove fluctuations of wavelength smaller than a certain
scale, which depends on (and increases with) the number of cooling 
steps.  Suppose, then,
that we observe an increase in abelianicity with cooling 
steps up to a certain maximum, followed by a decrease in abelianicity
as number of cooling steps continues to increase.  A reasonable
interpretation of such behavior would be that there are structures
in the vacuum, of some intermediate length scale, which are more abelian
in character than vacuum fluctuations at larger or smaller scales.
This "hill" of abelianicity (or dip in non-abelianicity), is in fact
the behavior we find.

\begin{figure}
\centerline{\hbox{\psfig{figure=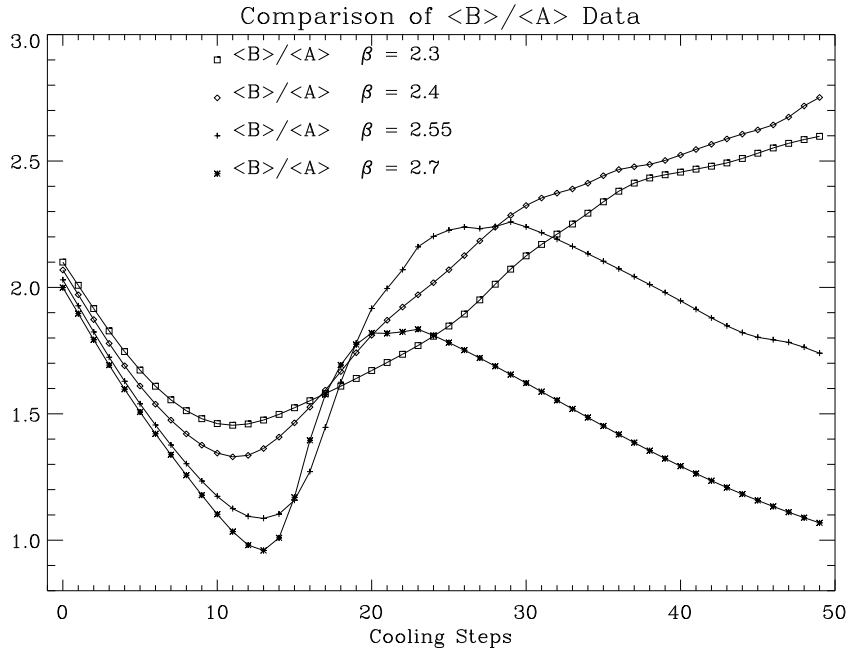,width=12.5cm,angle=0}}}
\caption[papfig1b]{Non-abelianicity $Q$ vs. Cooling Step, for several
values of $\beta$.}
\label{papfig1b}
\end{figure}

     Fig. 1 is a plot of the $Q$ observable vs. cooling step,
for $\beta=2.3,~2.4,~2.55$ and $2.7$ with $Q$ evaluated on $14,~17,~18$
and $25$ cooled configurations, respectively. In each case we worked on a 
$12^4$ lattice, thermalized for $3000$ iterations before applying the
cooling algorithm.  What is striking about this data
is the evident drop in $Q$, implying an increase in abelianicity,
up to $10-13$ cooling steps, which is followed by a rise in $Q$
(drop in abelianicity).  The effect seems to become more pronounced
as $\beta$ increases.  Again, the simplest interpretation is that
vacuum fluctuations at some intermediate length scale are more
abelian in character than vacuum fluctuations at smaller and at
somewhat larger length scales.
As cooling removes the higher frequency fluctuations, the contribution
of these "more abelian" fluctuations becomes relatively larger,
and $Q$ decreases.  As cooling proceeds, the "more abelian" fluctuations
are also removed, and $Q$ increases again.  

   It is interesting that the abelianicity begins to fall off after
$10$ cooling steps, because this is also where the plateau in Creutz
ratios vs. cooling step, noted by Campostrini et al. \cite{Pisa}, begins to
drop off to zero.  However, this simultaneous falloff of abelianicity
and Creutz ratio could be coincidental, particularly in light of
Teper's observation \cite{Teper} that the string tension should 
never disappear with cooling, providing one looks at 
sufficiently large loops.  

\begin{figure}
\centerline{\hbox{\psfig{figure=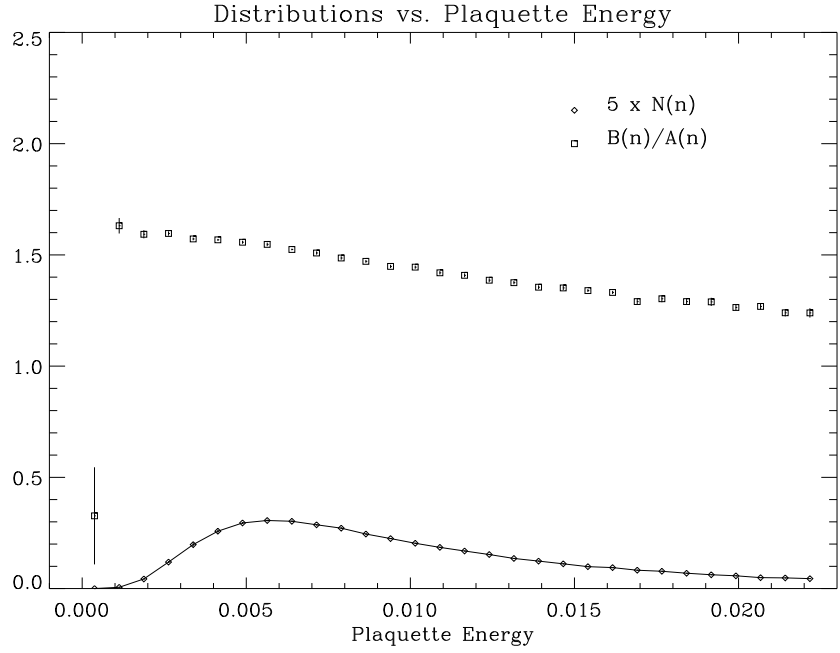,width=12.5cm,angle=0}}}
\caption[papfig2d]{Local non-abelianicity $Q(n)$ vs. Plaquette Energy after
$10$ cooling steps, at $\beta=2.55$.}
\label{papfig2d}
\end{figure}

\begin{figure}
\centerline{\hbox{\psfig{figure=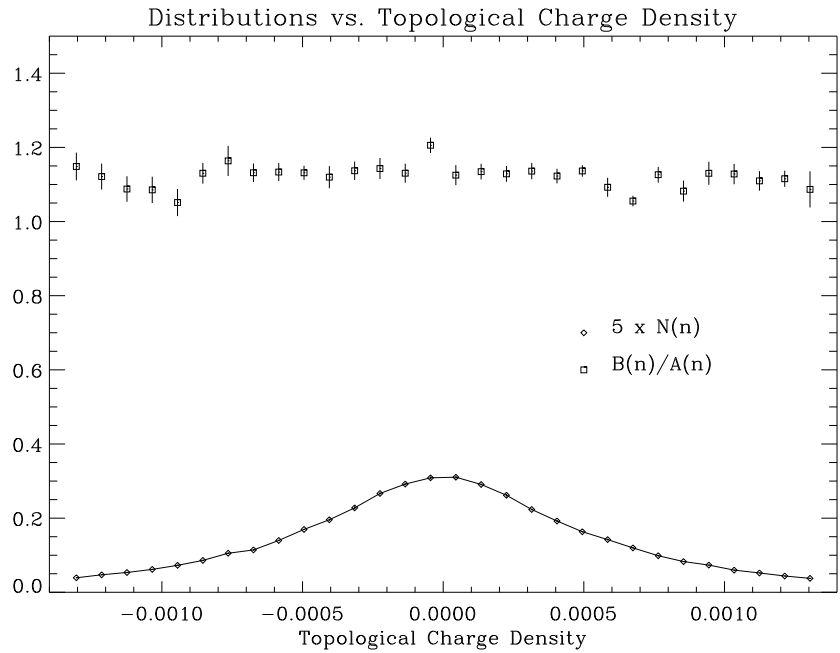,width=12.5cm,angle=0}}}
\caption[papfig3b]{Local non-abelianicity $Q(n)$ vs. Topological
Charge Density, after $10$ cooling steps, at $\beta=2.55$.}
\label{papfig3b}
\end{figure}

   The asymptotic behavior of $Q$ at large numbers of cooling steps is 
also of great interest,
but here we are rather hesistant in drawing conclusions from our
data.  In the first place, while the abelianicity at $\beta=2.3,~2.4$
steadily decreases, from $11$ cooling steps and beyond, the data
for $\beta=2.55,~2.7$ shows the opposite behavior: the abelianicity
reaches a minimum at about $23-25$ cooling steps, and then increases
again.  However, at the larger $\beta$ values a $12^4$ lattice is quite small,
and it is possible that beyond 25 cooling steps our observable is
mainly probing finite-size effects.

   The next question is whether the abelianicity is correlated with
either plaquette energy or topological charge.  We investigate this
issue at $\beta=2.55$ after $10$ cooling steps, which is near the
maximum average abelianicity.  The relevant data
is shown in Figs. 2 and 3, where 
the quantity $Q(n)$ in Fig. 2 is defined as follows: We first 
divide the range of plaquette energy into a
number of small intervals, indexed by $n$, and define $A(n)$ and $B(n)$ as 
\bea
  B(n) &=& -{1 \over N(n) V} \sum_x^{\mbox{bin~} n} {1 \over n_p (n_p - 1)} 
   \sum_{i>j} \sum_{m>n} \mbox{Tr} \{ [F_{ij}(x),F_{mn}(x)]^2 \}
\non \\
      A(n) &=& {1 \over n_p N(n) V} \sum_x^{\mbox{bin~} n} 
\sum_{i>j} \mbox{Tr}\{ F_{ij}^4 \}
\eea
where 
\beq
       N(n) = {1\over V} \sum_x^{\mbox{bin~} n} 1
\eeq
is the fraction of sites $x$ where the averaged plaquette value at the site
\beq
       S_p  =   {1\over n_p} \sum_{i>j} (1 - \oh 
     \mbox{Tr}[U_i(x) U_j(x+i) U^\dagger_i(x+j) U^\dagger_j(x)])
\eeq
is in the n-th interval.  Then
\beq
         Q(n) \equiv {<B(n)> \over <A(n)>}
\eeq

\noindent Each data point shown in Fig. 2 represents the data
from one interval; the x-component (plaquette energy) of the data point
is at the center of the interval, and the width of the interval is the
distance, along the x-axis, between neighboring data points.
Figure 3 is similar to Figure 2 , except that it is the
(naive) topological charge density
\beq
      {\cal T} =  -{1\over 32 \pi^2} \epsilon_{ijkl}
          \mbox{Tr}[U_{ij}(x) U_{kl}(x)]
\eeq
which is subdivided into intervals, where $U_{ij}(x)$ is the product
of link variables around a plaquette.

\begin{figure}
\centerline{\hbox{\psfig{figure=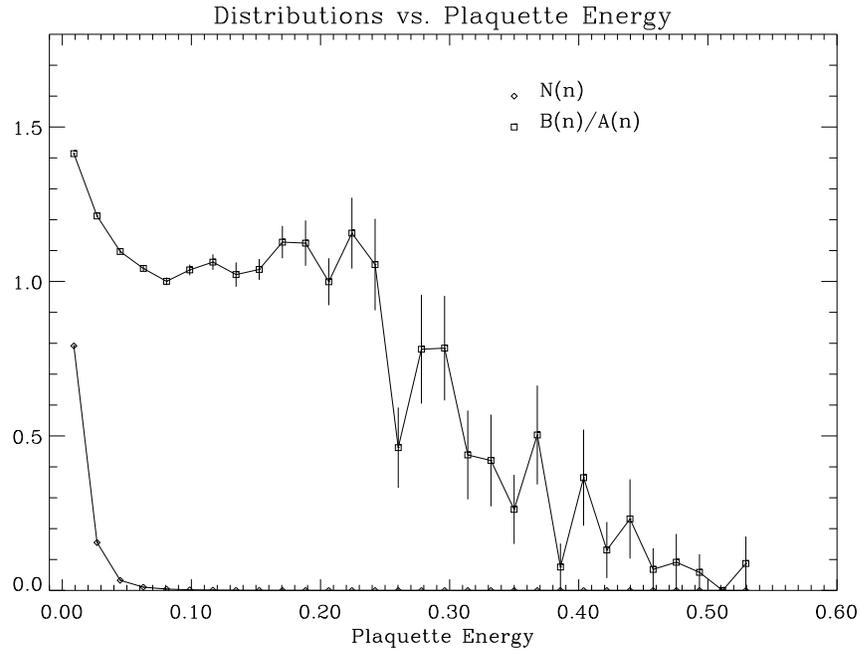,width=12.5cm,angle=0}}}
\caption[papfig4]{Same as Fig. 2, extended to larger plaquette energies.}
\label{papfig4}
\end{figure}

\begin{figure}
\centerline{\hbox{\psfig{figure=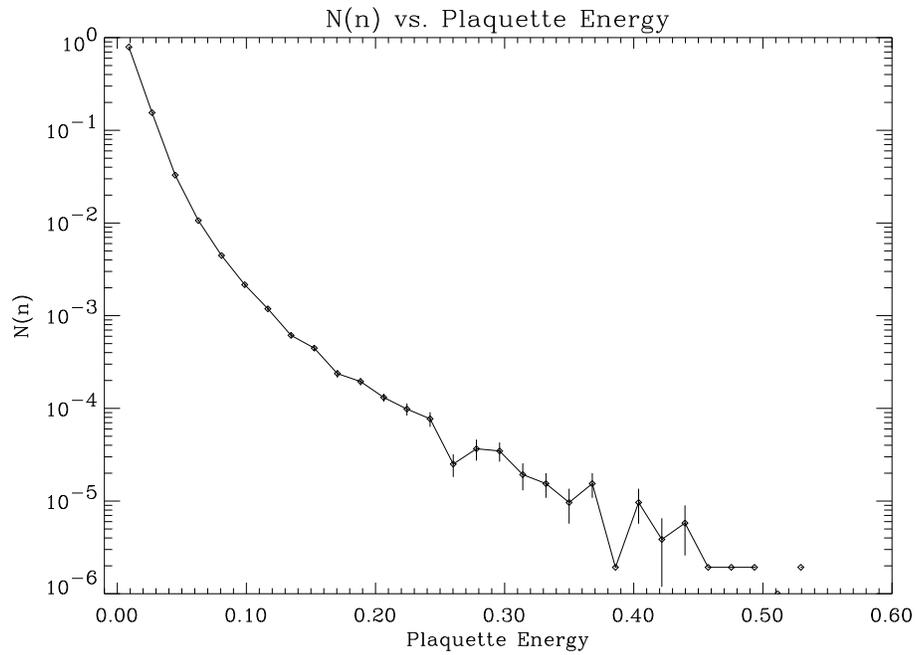,width=12.5cm,angle=0}}}
\caption[papfig6]{Fraction of sites $N(n)$ with plaquette energies in
the n-th bin, corresponding to data points in Figure 4.}
\label{papfig6}
\end{figure}

\begin{figure}
\centerline{\hbox{\psfig{figure=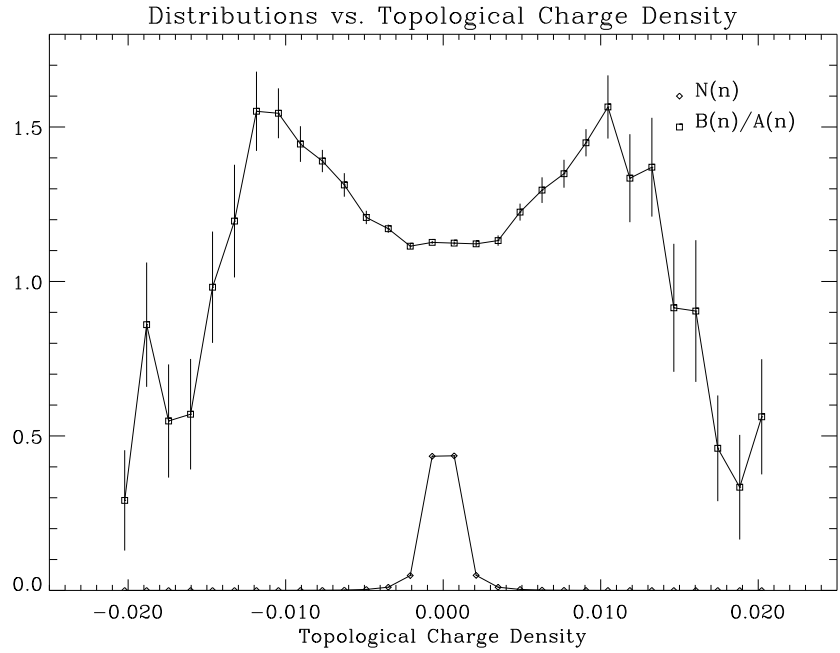,width=12.5cm,angle=0}}}
\caption[papfig6d]{Same as Fig. 3, extended to larger topological charge.}
\label{papfig6d}
\end{figure}

\begin{figure}
\centerline{\hbox{\psfig{figure=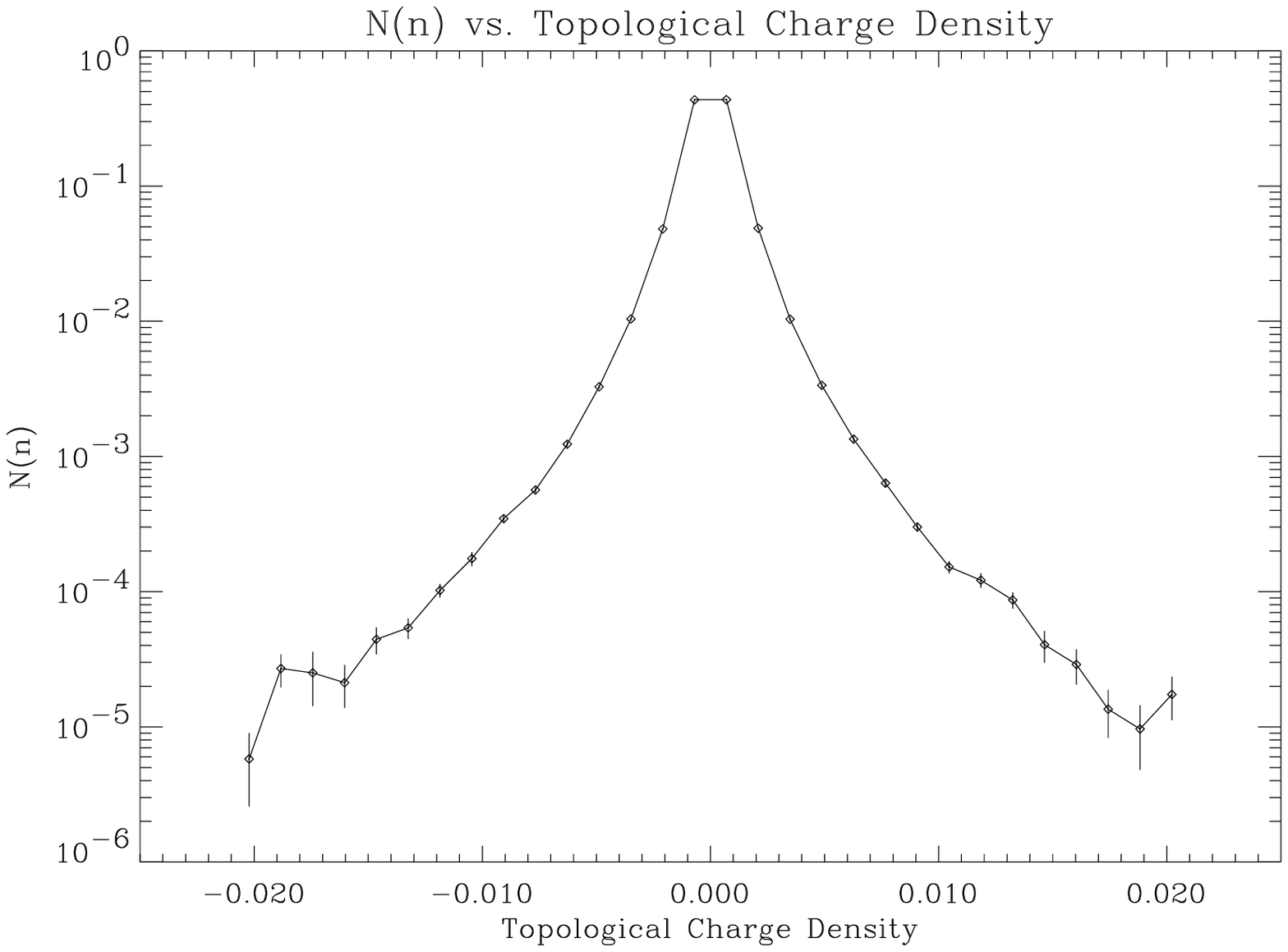,width=12.5cm,angle=0}}}
\caption[papfig7d]{Fraction of sites $N(n)$ with topological charge
density in the n-th bin, corresponding to data points in Figure 6.}
\label{papfig7d}
\end{figure}

   Figures 2 and 3 plot $Q(n)$ vs. plaquette energy $S_p$
and topological charge ${\cal T}$, respectively.  
We also display the average fraction 
$N(n)$ of sites with plaquette energy (Fig. 2) or topological
charge density (Fig. 3) in the n-th interval (or "bin")
associated with each data point.  The fraction $N(n)$ has been multiplied
by a factor of 5, to make the distribution more visible on the scale of the
graphs.  Apart from the very first (rather "noisy") data point in the
lowest plaquette energy interval, there    
does not seem to be a very strong correllation, after 10 cooling steps,
of abelianicity with either plaquette energy or topological charge in the
range shown. 

   The situation changes, however, if we include data which is off the
scale of Figures 2 and 3.  Figure 4 is another plot of $Q(n)$ vs. plaquette
energy at $\beta=2.55$ after $10$ cooling steps, with the scale of plaquette 
energy $S_p$ extended to a maximum of $0.6$ (the width of the
binning intervals is also increased).  It can be 
seen that as the plaquette energy increases beyond $0.25$, there 
is a steep increase in the abelianicity of the lattice field.  Fig. 5 
shows the fraction $N(n)$ of sites with averaged plaqette energies $S_p$ 
in the n-th bin.  It is evident that there are very few lattice
sites (on the order of $1$ in $10,000$) with $S_p > 0.25$.     
Similar behavior is seen for the abelianicity vs. topological charge
(Fig. 6), 
except that the abelianicity decreases at first, up to a minimum at
topological charge densities of magnitude $0.01$, after which there is again
a sharp increase in abelianicity.  As seen in Fig. 4-7, sites where
the abelianicity is far different from the average are very rare.
It is certainly intriguing that large plaquette energy and large topological
charge density are so strongly correlated with large abelianicity.
However, the rarity of sites with very large abelianicity means that
their physical importance is, as yet, uncertain.

   We conclude that there is some modest evidence that vacuum
fluctuations at an intermediate scale are more abelian in character
than fluctuations at smaller and at somewhat larger scales.  Conceivably,
this might indicate the presence of abelian domains.  The evidence
concerning the abelianicity of very long-wavelength fluctuations
is ambiguous, perhaps due to our relatively small lattice size. 
It would be interesting to study the asymptotic abelianicity of cooled
configurations on lattice sizes much larger than the $12^4$ volume used
here.

\vspace{33pt}

\noindent {\Large \bf Acknowledgements}

\bigskip

This work is supported in part by the U.S. Dept. of Energy, under
Grant No. DE-FG03-92ER40711, and also by the Director, Office of Energy
Research, Office of Basic Energy Sciences, of the U.S. Department of
Energy under Contract DE-AC03-76SF00098.


\end{document}